\documentstyle[12pt]{article}
\newcommand{\beq}{\begin{equation}}
\newcommand{\eeq}{\end{equation}}

\begin{document}
\def\lag{\langle}
\def\rag{\rangle}
\begin{titlepage}
\vspace{2.5cm}
\baselineskip 24pt
\begin{center}
\large\bf{Multicanonical Recursions$^{1}$}\\
\vspace{1.5cm}
\large{ Bernd A. Berg$^{2,3}$ }\\
\vspace{0.3cm}
(berg@hep.fsu.edu)
\end{center}
\vspace{3cm}
\begin{center}
{\bf Abstract}

The problem of calculating multicanonical parameters recursively
is discussed. I describe in detail a computational implementation
which has worked reasonably well in practice.
\end{center}

\vfill

\footnotetext[1]{{This research was partially funded by the
Department of Energy under contracts DE-FG05-87ER40319
{}~~~~and DE-FC05-85ER2500.}}
\footnotetext[2]{{Department of Physics, The Florida State University,
                      Tallahassee, FL~32306, USA. }}
\footnotetext[3]{{Supercomputer Computations Research Institute,
                      Tallahassee, FL~32306, USA.}}
\end{titlepage}
\baselineskip 24pt

\section{Introduction}

Recently Monte Carlo (MC) sampling with respect to unconventional ensembles
has received some attention [1--20]. In the multicanonical
ensemble \cite{our1,our3} one samples configurations such that exact
reconstruction of canonical expectation values becomes feasible for a
desired temperature range. This requires a broad energy distribution, and
leaves innovative freedom concerning the optimal shape \cite{Oxford}.
Considerable practical experience exists only for the uniform energy
distribution, where one samples such that:
\medskip

(a) The energy density is flat in a desired range
$$ P(E) = const ~~~{\rm for}~~~ E_{\min} \le E \le E_{\max} . \eqno(1) $$

(b) Each configuration of fixed energy $E$ appears with the same likelihood.
\medskip

It should be noted that condition (b) is non--trivial. A simple algorithm
\cite{RC} exists to achieve (a), but which give up (b). Exact connection
to the canonical ensemble is then lost. Such algorithms are interesting
for hard optimization problems, but unsuitable for canonical statistical
physics. The present paper focuses on achieving (a) and (b).

The average computer time $\tau$, measured in updates, which it takes to
proceed from $E_{\min}$ to $E_{\max} \ ~\underline{\rm and}$ back has been
named ``tunneling time'' \cite{our1,our2}. It should be noted that the
method overcomes free energy barriers actually not by a tunneling process,
but through moving along valleys, which are connected to the disordered
phase. Once an updating scheme is given, like standard Metropolis,
it is an interesting theoretical question to find the weight factors
which minimize the tunneling time. It is by no means clear that this
will be the uniform choice (1), on which the present paper is focused.

Multicanonical and related sampling has allowed considerable gains in
situations with "supercritical" slowing down. Such are:

\begin{description}

\item{(a)} First order transitions \cite{our1,our4}, for a recent review
           see \cite{Janke}.

\item{(b)} Systems with conflicting constraints, such as spin glasses
           \cite{our2,temp,BHC,Ker1} or proteins \cite{HO,HS}.

\end{description}

To achieve a flat energy distribution, the appropriate unnormalized weight
factor $w(E)$ is the inverse spectral density $w(E)=n^{-1}(E)$, just like
the weight factor for canonical MC simulations is the Boltzmann factor
$w^B(E)=\exp (-\beta E)$. Now, the spectral density is a--priori unknown.
Otherwise we would have solved the problem in the first place. Presumably,
reluctance about simulations with an a--priori unknown weight factor
is the main reason why the earlier umbrella sampling \cite{um} never
became popular in statistical physics.

For first order phase transitions the problem of the a--priori unknown
weight factor is rather elegantly overcome by means of finite size
scaling (FSS) methods \cite{our1,our4,Julich,BNB,Janke}. A sufficiently
accurate estimate is obtained by extrapolation from the already simulated
smaller lattices. The smallest lattices allow still for efficient
canonical simulations.

For systems with conflicting constraints the situation is less
satisfactory. For instance for spin glasses one has to perform the
additional average over quenched random variables (which are the exchange
coupling constants). Different choices of these random variables define
different realizations of the same system. For the Edward--Anderson
Ising spin glass it turned out \cite{our2,BHC}
that, even for identical lattice sizes, different realizations need
different weight factors. Each system requires a new estimate of the
weight factors with no a--priori information available. To achieve
this, a recursion$^{23}$
was introduced by Celik and the author \cite{our2}.
However, details of the recursion (see section 3) may need considerable
attention by hand. This attention is possible when only a few lattices
are simulated, but impractical when hundreds or even thousands of
different realizations have to be handled. This renders it inconvenient
for more complicated situations, like the 3$d$ Edwards--Anderson
Ising (EAI) spin glass.

Consequently, the recursion actually used in Ref.\cite{BHC}, where
multicanonical simulations were performed for more then 1,500 different
realizations of the EAI model, differed from the one described in
\cite{our2}.  The main purpose of this article is to describe this
particular approach. In each recursion step the statistical information
from all previous runs is used directly for estimating the multicanonical
parameters as well as for noise reduction. Further, the recursion turned
out to be robust.
Little attention by "hand" was needed. However, no claim is made that it
is in any sense optimal (actually the author is considering various
improvements). It is supposed to be a reasonable starting point to
provide a running code quickly.

The paper is organized as follows: In section 2 generalized Ising models
and related preliminaries are introduced. Mainly for pedagogical reasons
I focus on them for examples of this paper. It is clear that generalization
to other
systems is straightforward, although possibly tedious for continuous systems.
In section 3 I introduce the multicanonical method, and discuss the
recursions given in the literature \cite{our2,Lee}. Section 4 describes the
recursion which I invented for the simulations of \cite{BHC}, and section~5
illustrates its performance. Summary and conclusions follow. The appendix
gives and explains a corresponding program listing.
\hfill\break

\section{Generalized Ising Models}

Let us consider a $d$--dimensional hypercubic
lattice of volume $V=N=L^d$ with periodic boundary conditions.
Spins $s_i=\pm 1$ are located at the $N$ sites, and exchange
interactions $J_{ik}=\pm 1$  at the $dN$ links of the lattice.
The energy of generalized Ising models is given by
$$ E = - \sum_{ik} J_{ik} s_i s_j ,               \eqno(2)$$
where the sum is over the nearest neighbors. For
$J_{ik} \equiv 1$ the standard Ising ferromagnet (IF) is recovered. When
the $J_{ik}$ are quenched random variables, one obtains the EAI spin glass.
I confine the subsequent discussion to these two situations, although
there are other cases of interest \cite{FFI}. Let us further restrict the
EAI spin glass to the situation $\sum_{<ik>} J_{ik} = 0$.

The partition function may be written as
$$  Z (\beta ) = \sum_E n(E) e^{-\beta E} , \eqno(3) $$
where $n(E)$ is the spectral density \cite{KH}, more precisely
the number of configurations (or states) with energy $E$. As the
system has $2^N$ different states, this implies the normalization
$$ \sum_E n(E) = 2^N . \eqno(4)$$
The lowest possible energy is $-dN$, reached when each link contributes
$J_{ik}s_is_k=1$. For the IF this is achieved with either
all spins up $(+1)$ or all spins down $(-1)$. For a generic configuration
the possible energy increments under the flip of a single spin are
$$ \triangle E = 0, \pm 4, ..., \pm 4d . \eqno(5)$$
Consequently $n(E)$ may take non--zero values for
$$ E = -dN, -dN+4, ..., 0, ..., dN-4, dN . \eqno(6)$$
For instance for the IF $n(-dN)=2$, $n(-dN+4)=0$, ...,
and $n(-dN+4d)=N$. For a typical EAI spin glass configuration the
groundstate energy $E_{\min}$ is considerably larger than $-dN$.
\hfill\break

\section{Multicanonical Sampling}

In the pedagogical review \cite{our3} I emphasized that the inverse
spectral density is the appropriate weight factor to obtain a
flat energy density
$$ w(E) = n^{-1} (E) = e^{-\beta (E) E + \alpha (E)} . \eqno(7) $$
Here $\beta (E), \alpha (E)$ is the multicanonical parameterization
\cite {our1,our2,Bau}. Its rationale is to relate to the temperature. It
should be noted that MC calculations are insensitive to an over--all
independent factor, {\it i.e.} against replacing $w(E)$ by $c\, w(E)$.
In the following I will exploit this property
from time to time, and not trace back the corresponding multiplicative
or additive constants. If necessary they may be obtained by introducing
a convenient normalization. The spectral density may be written as
$$ n(E) = e^{S(E)} , \eqno(8) $$
where $S(E)$ is the microcanonical entropy \cite{KH}. The thermodynamical
relation for the inverse temperature ($\beta = T^{-1}$, where my
Boltzmann constant convention is $k=1$) is
$$ \beta = {\partial S \over \partial E} . \eqno(9)$$
For models with discrete energy values this may be translated into
$$ \beta (E) = {S(E+\epsilon)-S(E) \over \epsilon} , \eqno(10)$$\
where $\epsilon$ is the smallest possible energy increment such that
$n(E+\epsilon)$ and $n(E)$ are both non-zero. {\it I.e.} typically we
have $\epsilon=4$ for the model of section 2 (special care is needed for
the IF close to its groundstate). Note that equation (10) is
in part convention. Other valid options would be
$\beta (E) = [S(E)-S(E-\epsilon)]/\epsilon$ or
$\beta (E) = [S(E+\epsilon)-S(E-\epsilon)]/(2\epsilon)$. For consistency
with \cite{our2,our3} I stay with (10).

Once $\beta(E)$ is given, $\alpha(E)$ may be determined recursively.
The equality of $e^{-S(E)}$ and $e^{-\beta(E)E+\alpha(E)}$ implies
$$ S(E)-S(E-\epsilon) = \beta(E)E - \beta(E-\epsilon)(E-\epsilon)
   -\alpha(E)+\alpha(E-\epsilon) . $$
Using (10) to eliminate the term $\epsilon\beta(E-\epsilon)$, we find
for $\alpha(E)$ the recursion relation
$$ \alpha (E-\epsilon) = \alpha (E) + [\beta(E-\epsilon)-\beta(E)] E,
{}~~~\alpha(E_{\max}) =0 . \eqno(11) $$
Here $\alpha(E_{\max})=0$ is a choice of the over--all
multiplicative constant, needed to start off the recursion.

To perform a multicanonical simulation, we do not need to know the exact
weight factor (7). Instead, a working estimate $\overline{w} (E)$ of
$w(E)$ is sufficient, such that the sampled energy histogram $H(E)$ is
approximately flat in the desired energy range (1). In the subsequent
discussion I use the notation $\overline{n}(E)$, $\overline{S}(E)$,
$\overline{\beta}(E)$, $\overline{\alpha}(E)$ for estimators of the
corresponding quantities $n(E)$, $S(E)$, $\beta (E)$ and $\alpha (E)$.

The technical feasibility of multicanonical sampling depends on the
existence of efficient methods to obtain an acceptable estimate
$\overline{w}(E)$. Computational resources and concurrent numerical
options determine how much computer time one will consider acceptable
for the calculation of $\overline{w}(E)$. Typically it should be less
or at most of the order of the CPU time spent on the subsequent
multicanonical sampling (with then fixed parameters). It seems that
different workers in the field have tried various approaches. I am
only familiar with two of them.

\begin{description}

\item{(a)} Approaches which work in one or two steps \cite{our1,our4,BNB}.
Employing FSS a reasonable good approximation $\overline{w}^{(1)} (E)$
is obtained by extrapolation from previously simulated, smaller
lattices. With ${\overline w}^{(1)} (E)$ a first multicanonical simulation
is carried out. Its results give an improved estimate
$\overline{w}^{(2)} (E)$ with
which additional simulations may be done. This approach works well for first
order phase transitions, but failed badly for some disordered systems.

\item{b)} Recursive calculations $\overline{w}^n (E) \to
\overline{ w}^{n+1} (E)$ have been employed. They are subject of the
following subsection.

\end{description}

\subsection{Recursive multicanonical calculations}

Let $H^n (E)$ be the unnormalized histogram obtained from a (short)
multicanonical simulation with ${\overline w}^n (E)$. At energy values
for which $H^n (E)$ is $\underline{\rm reliable}$, the new estimate is
$$ \overline{w}^{n+1} (E) = {\overline{w}^n (E) \over H^n (E)} . \eqno(12)$$
Clearly (12) fails for energy values for which $H^n(E)=0$, and also
values like $H^n(E)=1$ or 2 are of course statistically unreliable.
Worse, even large values like $H^n(E)=10^6$ may still not give reliable
estimates. Namely, situations can be encountered where the integrated
autocorrelation time is of the same order of magnitude
or even larger. Before I come to a more thorough discussion of this
problem, I would like to discuss two approaches in the literature.

To be definite, let us assume that the starting point for the recursion
is
$$ \overline{w}^0 (E) \equiv 1 . \eqno(13) $$
In general this is a reasonable choice, which will allow us to recover
the normalization (4) when desired. For some practical applications other
choices, like a canonical simulation at a certain temperature, may be
more convenient.

In the paper by Celik and myself \cite{our2} equation (12) was stated in
the multicanonical notation (7). It reads then (note
$\epsilon =4$ in Ref.\cite{our2})
$$ \overline{\beta}^{\, n+1}(E) = \overline{\beta}^{\, n} (E) +
\epsilon^{-1} \ln [ H^n(E+\epsilon) / H^n(E) ] .  \eqno(14a) $$
The function $\overline{\alpha}^{n+1}(E)$ is then determined by
equation (11). In addition to (14a) specific rules were given about
how to exclude unreliable histogram entries. Namely,
$$ \overline{\beta}^{\, n+1}(E) = \cases{
\overline{\beta}^{\, n}(E) ~~{\rm for}~~ E\ge E^n_{\rm median};\cr
\overline{\beta}^{\, n+1}(E^n_{\rm cut-off}) ~~{\rm for}~~
E<E^n_{\rm cut-off} .}   \eqno(14b)$$
Here $E^n_{\rm median}$ is the median of the $n^{th}$ energy distribution,
and $E^n_{\rm cut-off} < E^n_{\rm median}$ is an energy cut--off, such
that in simulation $n$ the temperature is kept constant for
$E<E^n_{\rm cut-off}$. Further, note that the starting condition (13)
becomes $\overline{\beta}^{\, 0}(E)\equiv 0$,
$\overline{\alpha}^0(E)\equiv 0$.

Lee \cite{Lee} states his recursion in two parts:
$$ \overline{S}^{n+1} (E) = \overline{S}^n (E) +
   \ln H^n(E) ~~~{\rm for }~~~ H^n(E) \ge 1, \eqno(15a) $$
and
$$ \overline{S}^{n+1} (E) = \overline{S}^n (E)
{}~~~{\rm for }~~~ H^n(E) = 0, \eqno(15b) $$

The first part is obviously equation (12), as follows from
$\overline{w}^n(E)=\exp[-\overline{S}^n(E)]$. The identity \cite{our5}
of (15a) and (14a) follows from (10). Obviously (15a) is a
convenient intermediate step to derive (14a). The second part (15b)
is a specific prescription about how to handle $H^n(E)=0$. The other
unreliable $H^n(E)$ are included into the recursion (12).
Let us note the following:

\begin{description}

\item{(a)} Besides from minor notational differences, it is uniquely
determined how to handle the reliable part of the data. One should
note that the equivalent equations (12), (14a) and (15a) are all
non--local in the sense that ultimately histogram entries over the
entire sampled range will determine the transitions amplitudes from
one energy to the next. It may be a little surprising that equation (14a)
looks less local than equation (12) or (15a). This is entirely irrelevant,
because the weight factors are only auxiliary quantities to
determine (for instance by detailed balance) the
$\underline{\rm decisive}$ transition probabilities $w[E\to E']$.
The transition probabilities relate $\underline{\rm different}$ energies.
$W = \left(w[E\to E']\right)$ forms a (sparse) matrix, and its
eigenvector with eigenvalue one is finally supposed to become the
spectral density, {\it i.e.} determines the weight factors. This
diagonalization (implicitly carried out by the MC simulation) is a
non--local process. This non--locality may induce certain instabilities.
For instance, if inaccurate weight factors somewhere in the energy
range create a region of attraction, all CPU time in the next run
may be wasted on iterating on an irrelevant energy region.

\item{(b)} As the recursion (12) and (15a) stand, the statistical
accuracy of estimate $n+1$ is entirely determined by MC simulation $n$.
With increasing $n$ the covered energy
range gets larger and larger. One needs longer and longer simulations
just to regain the previously reached statistical accuracy (on the
appropriate energy subrange). It is possible, but tedious, to combine
the statistics of simulations $n$, $n-1$, ..., 1, 0. The gain is not
quite as dramatic as one may superficially expect, because simulation
$n+1$ does still explore the entire energy range, just for the purpose
to explore an additional increment of the desired energy range.

\item{(c)} Note that the median rule of (14b) freezes estimates on
some part of the already covered energy range. One should improve
on it by using subsequent statistics when available. In \cite{our2} it
was suggested to combine the median rule with upper bounds on the energy,
such that the energy range gets reasonably restricted. However, it is
then difficult to ensure ergodicity.

\item{(d)} A central difficulty of the recursion is the handling of
energy regions for which reliable statistical information is not yet
available. I elaborate on this now.
\end{description}

Lee's proposal (15b) looks attractive because of its simplicity. It
works for the very small systems considered in his paper, but for
many realistic situations it will lead to an unacceptable slowing down.
The reason is that (15b) is equivalent to simulating with a
constant weight factor (7). Now, at low temperatures one typically
encounters
$$ n(E-\epsilon)/n(E) \sim V^{-1} . \eqno(16) $$
Therefore, for a not yet covered energy range $E\le E_0$ one will need of
order $V$ attempts just to achieve once the transition
$E_0 \to E_0 - \epsilon$.

The rule $\overline{\beta}^{\, n+1} =
\overline{\beta}^{\, n+1} (E_{\rm cut-off})$ for $E<E_{\rm cut-off}$ from
(14b) achieves a far better performance for this situation. Assume that
$\beta (E)$ is monotonically increasing towards lower energies (exceptions
are first order phase transitions). A
$\underline{\rm canonical}$ simulation with $\beta^{n+1} (E_{\rm cut-off})$
will have its maximum energy density at $E=E_{\rm cut-off}$, because its
first derivative with respect to the energy is zero there.
The width of its energy distribution is of order $\sqrt{V}$.
Consequently, there will be no weight factor problem associated with
proceeding towards lower energies. In practice one has to use estimators
$\overline{\beta}^{\, n+1}(E)$. One would like to chose $E_{\rm cut-off}$
as low as possible, but one encounters noise problem when the cut-off energy
is shifted too far towards the edge of the reliably covered energy range. With
some experience a good ``pick'' for $E_{\rm cut-off}$ can be achieved by
just inspecting the function $\overline{\beta}^{\, n+1}(E)$. Alternatively,
one may use a fit $\overline{\beta}^{\, n+1}_{\max}$ from several energy
values instead of $\overline{\beta}^{\, n+1}(E_{\rm cut-off})$, or even fit
the continuation of the entire function $\overline{\beta}^{\, n+1}(E)$ for
$E<E_{\rm cut-off}$ (with the penalty of spurious instabilities). In any
case, in energy regions where (16) holds, one expects a performance
increase by at least a volume factor over using (15b).
On the other hand, it is precisely this part of the recursion (14)
which required annoying attention by hand. This experience can, of course,
note rule out the possible existence of some more perfect fitting
procedure, to estimate $\overline{\beta}^{\, n+1}(E)$ towards lower
energies.

How the recursion (14) slows down with volume depends thus on the
details of its implementation. Typically, one has to cover
a macroscopic energy range, {\it i.e.} $E_{\max}-E_{\min}\sim V$. The
optimal slowing down of a single multicanonical simulation on this range
is $\sim V^2$, corresponding to a random walk in the energy \cite{our1}.
Of order $V^{0.5}$ simulations are needed to iterate from an initial
canonical distribution up to covering the entire energy range
multicanonically.
This leads to an optimal slowing down $\sim V^{2.5}$ for the recursion. That
this is not an overestimate follows from the fact that the slowing down of
a multicanonical simulation on half the energy range still scales with
$V^2$, and it still takes of order $V^{0.5}$ simulations to iterate from
half the range to the full range.
\hfill\break

\section{Accumulative Recursion}

I now introduce a recursion which calculates
$\overline{\beta}^{\, n+1}(E)$
on the basis of the statistics accumulated in all previous runs
$n, n-1, ..., 1$. For this purpose let us first re--write (14a) as
$$ \overline{\beta}^{\, n+1} (E) = \epsilon^{-1}
\ln [ H^n(E+\epsilon) / H^n_{\beta} (E) ] ,  \eqno(17) $$
where
$$ H^n_{\beta}(E) = H^n(E)
e^{-i\overline{\beta}^{\, n}(E)\epsilon} . \eqno(18) $$
Equation (17) still holds when $H^n(E)$ and $H^n_{\beta}(E)$ are
replaced by non--zero linear combinations $\hat{H}^n(E)$ and
$\hat{H}^n_{\beta}(E)$:
$$ \hat{H}^n (E) = \sum_{m=0}^n W^m (E) H^m (E) , \eqno(19a) $$
$$ \hat{H}^n_{\beta} (E) =
	   \sum_{m=0}^n W^m (E) H^m_{\beta} (E) . \eqno(19b) $$
The accumulated statistics can be presented by suitable choice of the
weight factors $W^m (E)$. The optimal choice is not clear,
as it may depend non-trivially on the dynamics. In practice
$$ W^m (E) = {\min [H^m(E+\epsilon),H^m(E)] \over
                         \max [H^m(E+\epsilon),H^m(E)]} \eqno(20)$$
has worked well. It relies on the conservative assumption that each
contribution to the estimate
$$ \overline{\beta}^{\, n+1} (E) = \epsilon^{-1}
\ln [ \hat{H}^n(E+\epsilon) / \hat{H}^n_{\beta} (E) ]  \eqno(21) $$
will be as good as its weakest part. This equation is supplemented by
$$\overline{\beta}^{\,n+1}(E) =
\overline{\beta}^{\,n+1}(E+\epsilon) \eqno(22)$$
for the case that either $\hat{H}^n(E+\epsilon)$ or
$\hat{H}^n_{\beta}(E)$ has insufficient statistics. To provide some
feeling for the estimator (21) let me discuss two special cases.

\begin{description}
\item{(a)} When the desired, flat distribution is already reached, the
weight factors (20) equal 1 up to statistical fluctuations. Let us
ignore fluctuations for the moment. Then $\hat{H}^{n-1}(E+\epsilon) =
\hat{H}^{n-1}(E)$ holds before the $n^{th}$ run, which uses
$\overline{\beta}^{\, n}(E)$ as defined by equation (21). In the
$n^{th}$ recursion $H^n(E+\epsilon)=H^n(E)$ is obtained by assumption.
This leads to $\hat{H}^n(E+\epsilon)=\hat{H}^{n-1}+H^n(E+\epsilon)$ and
$\hat{H}^n_{\beta}(E)=\hat{H}^{n-1}_{\beta}(E)+H^n(E)\, \exp(-
\overline{\beta}^{\, n}\epsilon )$. Equations (19), (21) yield
$\overline{\beta}^{\, n+1} (E) = \overline{\beta}^{\, n} (E)$, {\it
i.e.} the $\beta (E)$ function is a fixed point when the sampled
distribution is flat.

\item{(b)} Consider the first recursion, carried out with
$\overline{\beta}^{\, 0}(E)\equiv 0$. The sampling results will be
$H^0(E+\epsilon)/ H^0(E) = n(E+\epsilon)/n(E)$, again up to
statistical fluctuations. Recursion (21) yields
$\overline{\beta}^{\, 1}(E)=\epsilon^{-1}\ln [n(E+\epsilon)/n(E)]$,
which is already the final multicanonical answer due to the fact
that we have neglected statistical fluctuations. Quite generally
it can be shown that the desired multicanonical function $\beta(E)$
is an attractive fixed point of the recursion.
\end{description}

In practice there may be severe statistical fluctuations due to only
few, correlated entries in $H^n(E+\epsilon)$, $H^n(E)$ or both. If
the number of entries in both arrays is small, but approximately
equal $(W^n(E)\approx 1)$, equations (19) guarantee that increase
from $\hat{H}^{n-1} \to \hat{H}^n$ is in proportion the the generated
statistics (assuming similar autocorrelation time in runs $n-1$, $n-2$,
...). If the number of entries is only small in either
$H^n(E+\epsilon)$ or $H^n(E)$, the weight factor (20) correct for
the asymmetry. The larger statistics is reduced to the smaller one,
and the smaller even more suppressed. As the ratio
$\hat{H}^n(E+\epsilon)/\hat{H}(E)$ determines the estimate
$\overline{\beta}^{\,n}(E)$, it is clear that a large statistical
fluctuations in either the numerator or the denominator is sufficient
to destroy the entire estimate. The weight factor prevents this.

The obvious advantage of equation (21) over recursion versions of
section~3 is that the accumulative statistics of all runs is used
to reduce statistical fluctuations. In \cite{BHC} we have not
supplemented the present recursion by a median restrictions of the
type (14b), although this might lead to further improvements. Without
such restrictions, typically the recursion leads quickly to rather high
$\overline{\beta}$ values, and works its way back from the corresponding
low energy values through the entire energy range. Occasionally this
has led to ``hang--up'' situations, for which a simple ``retreat''
strategy has turned out to be sufficient.
For the case of generalized Ising model, the appendix gives and
explains an actual program listing, which was used
for the numerical illustrations of the next section.
A generalization of my recursion to non--flat distributions, like for
instance those proposed in \cite{Oxford} would be straightforward.
\hfill\break

\section{Numerical Tests}

I confine myself to reporting results for the 3d IF and
the 3d EAI spin glass. Similar tests have been performed for the 2d
IF and are in progress for the 2d EAI spin glass as
well as for fully frustrated Ising models \cite{Weaver}. To keep
the relation to the program listing in the appendix close, I shall
use
$$ I_A = {1\over 4}\, (-E + dN), ~~~({\rm with}~~ N=L^d) \eqno(23) $$
instead of the energy, defined by (2). The rationale of $I_A$ is
its range:
$$ I_A = 0,\, 1,\, 2,\, ...,\, dN/2 \eqno(24) $$
in typical increments of 1. For comparison, we had
$-dN \le E \le dN$ in typical increments of 4. Consequently,
for the purposes of programing $I_A$ is far more convenient.
Functions of $E$ are now interpreted as functions of $I_A$ in the
obvious way, {\it i.e.} $\beta(E)=\beta[E(I_A)] \to \beta(I_A)$,
and so on.

\subsection{Three dimensional Ising ferromagnet}

The first few terms of the low temperature expansion on a finite
(but sufficiently large) lattice collected in table~1. The present
computer program is unsuitable to cope with $n(I_A)=0$ for
$I_A=(3N/2)-1,\, (3N/2)-2$ and $(3N/2)-4$. I just bypass$^{28}$
the problem by restricting the updating to the range
$I_A \le N_{\max} = (3N/2)-5$. Proposals with $I_A > N_{\max}$
are simply rejected.

We want to calculate multicanonical parameters for the temperature
range infinity down to zero. Simulations with $\beta \equiv 0$ are
peaked around $I_A=N_{\min}=3N/4$. We therefore fix the function
$\beta$ to $\beta(I_A)=0$ for $I_A \le N_{\min}$, and never change
it there. For $I_A>N_{\min}$ we perform the multicanonical recursion
of section~4. The covered range of lattices was $4\le L\le 16$.
In a first set of runs the recursion was applied until
the system tunneled at least 60 times. The (expected) experience from
these runs is that the recursion remained stable after the first
tunneling. The tunneling time $\overline{tau}$ is then measured
${\rm \underline{after}}$ the first tunneling has occurred, while
continuing to update the parameters.
Table~2 collects the measured tunneling times $\overline{\tau}$,
and states on how many
tunneling events $n_{\tau}$ the estimates actually rely.

By $\tau_0$ I denote the time (as always in updates) it takes until
the first tunneling has taken place. This is essentially the time our
recursion needs to provide a reliable estimate of the multicanonical
parameters, and it will therefore be called {\it recursion time} in
the following. Two estimates, $\overline{\tau}_0^a$ and
$\overline{\tau}_0^b$,
are given in table~2. They differ by the number of sweeps performed
before the multicanonical parameters are updated ({\it i.e.}
the subroutine {\tt UPMUCA}
of the appendix is called). A sweep is defined by updating $N=L^d$
spins. For $\overline{\tau}_0^{\,a}$
{\tt UPMUCA} was called every 120 sweeps, whereas
for $\overline{\tau}_0^{\,b}$ it was called every $N$ sweeps.
Respectively, this amounts to letting the numbers of intermediate
updates grow in proportion to the volume and to the volume squared.
Within the (still large) statistical errors there is no
difference noticeable.

The values $n_{\tau_0^a}$ and $n_{\tau_0^b}$ are the numbers of
$\beta(I_A)\equiv 0$ re--starts on which the respective estimates
rely. As the average CPU time needed per recursion is substantially
higher than the average tunneling time $\tau$, I have limited the
$\tau_0$ analysis to $L\le 12$. The given error bars are somewhat
unreliable as the obtained distributions have
long tails towards large $\tau_0$ values. A large
statistics is needed to get into the region where the central
limit theorem provides a good approximation. My typical number
of $n_{\tau_0}=126$ events is a bit at the low edge. Figure~1
employs a log--scale for $\tau_0$ to show the histograms for
$\tau_0^b$. The distributions for the tunneling times $\tau$
themselves, are more reasonably, Poisson like, behaved.

Figure~2 shows the increase of $\tau$ and $\tau_0^b$ with volume
on a log--log scale. The straight lines correspond to the fits
$\tau=cV^{\delta}$ and $\tau_0^b=c_0V^{\delta_0}$. The results
for the fit parameters are
$$ \ln (c) = -0.53 \pm 0.16, ~~ \delta = 2.249 \pm 0.021,
{}~~ (Q = 0.18)  \eqno(25) $$
and
$$ \ln (c_0) = -1.24 \pm 0.17, ~~ \delta_0 = 2.931 \pm 0.023,
{}~~ (Q = 0.70) , \eqno(26) $$
where $Q$ is the goodness of fit \cite{Recipes}. It should be
remembered that the lower bounds are $\delta = 2$ \cite{our1}
and $\delta_0 = 2.5$ (see section~3).

To demonstrate that after a few
tunneling events the multicanonical parameters are indeed already
useful, I have also measured a tunneling time $\tau_1$, obtained
by fixing  the multicanonical parameters after the first four
tunneling events. Table~2 contain also the corresponding estimates
$\overline{\tau}_1$. Within the statistical errors, there is no
difference with the estimates $\overline{\tau}$.

\subsection{Three dimensional Edwards--Anderson Ising spin glass}

First I present some results  from the extensive investigation \cite{BHC},
which are not contained in this reference. For fixed
lattice size $L$ tunneling times $\tau$ are found to vary greatly for
different $J_{ik}$ realizations. For each lattice size figure~3 connects
the tunneling times, sorted in decreasing order. For $L=4-8$ there are
512 different realizations per lattice. For $L=12$ there are only seven
realizations, depicted at $64 (i-1),\ (i=1,...,7)$. The lines are drawn
to guide the eyes. Figure~4 depicts histograms for the $L=4-8$
tunneling times. In both figures a logarithmic scale is used for $\tau$.
The worst realizations have dramatically larger tunneling times than
{\it typical} ones, defined by the median value $\tau_{0.5}$. This leads
to large differences between the mean value $\overline{\tau}$, which
determines the needed computer time, and the median value
$\overline{\tau}_{0.5}$.
These values are collected in table~3. With increasing lattice
size the discrepancy between mean and median increases dramatically
(the $L=12$ data have to be considered unreliable for this purpose).
This lack of self--averaging of the spin glass with respect to the
multicanonical tunneling time comes somewhat surprising, and remains
to be better understood. Also collected in the table are the smallest
$\tau_{0.0}$ and largest $\tau_{1.0}$ tunneling time, found on the
investigated realizations.

For typical spin glass realizations, {\it i.e.} the realizations
corresponding to the median $\tau_{0.5}$ tunneling times of table~3,
I have performed the same analysis as for the 3d IF in the previous
subsection. The results are collected in table~4. An interesting and
unexpected result is that I find $\overline{\tau}_1$ systematically
smaller than $\overline{\tau}$, {\it i.e.} further applications of
the recursion relation make the tunneling worse. My tentative
interpretation is that the flat distribution is not optimal. Due
to statistical fluctuations, one can then imagine that immediately
after one of the first few tunneling events the generated
multicanonical parameters are positively correlated towards a more
optimal choice. A more detailed future analysis may be desireable.

As before, the recursion times $\overline{\tau}^a_0$ and
$\overline{\tau}^b_0$ are practically identical. However, a second
unexpected result is that now the recursion times take the same order
of magnitude as the tunneling times, whereas for the 3d IF the
recursion times were considerably large than the tunneling times.

It has to be remarked that the $L=12$ results are not in line with
the subsequent estimates from lattices of size $L=4$, 6 and 8. (a)~The
$\overline{\tau}_1$ estimate is considerably higher than expected. The
reason is likely that the typical realization picked is not typical.
A reliable estimate of $\tau_{0.5}$ is practically impossible due to
the small number of only seven $L=12$ realizations investigated in
\cite{BHC}. (b) The $\overline{\tau}^b_0$ value, given in brackets, is
much smaller than expected. However, the number is given in brackets as
it is not an estimate of said quantity. Altogether twelve attempts were
made, to determine multicanonical parameters by means of the recursion.
Of those, two did not lead to a single tunneling event within the
maximally allowed CPU time corresponding to approximately 11.2E08
updates. These two attempts are (cannot be) not included in the given
average values. This behavior illustrates that one should perform
several independent starts, when applying the recursion to difficult
situations.

Using only the estimates from $L=4, 6$ and 8, the subsequent
my results are obtained from straight line fits to the equations
$\tau =cV^{\delta}$, $\tau_1 =c_1V^{\delta_1}$ and
$\tau_0^b=c_0V^{\delta_0}$:
$$ \ln (c) = -3.04 \pm 0.29, ~~\delta = 3.24 \pm 0.06,
{}~~(Q=0.40), \eqno(27a)$$
$$ \ln (c_1) = -3.61 \pm 0.24, ~~\delta = 3.28 \pm 0.05,
{}~~(Q=0.39), \eqno(27b)$$
and
$$ \ln (c_0) = -2.23 \pm 0.24, ~~\delta = 3.09 \pm 0.04,
{}~~(Q=0.78), \eqno(28)$$
Here, as well as in the previous section, the routine GFIT
from \cite{Recipes} gives results perfectly compatible with the
linear fit results. A figure corresponding to (27) and (28) looks
similar to figure~2, but is not very instructive as all three
fits lines are almost on top of one another.
The exponent $\delta$ is smaller than the one reported in \cite{BHC}.
The reason is that it is differently defined. In \cite{BHC} the
tunneling time was averaged over all realization, whereas here I have
picked single, typical realizations. There is evidence that for the
worst realizations the tunneling time slows down exponentially with $L$.
This spoils the power law fit for the average over all realizations.
\hfill\break

\section{Summary and Conclusions}

For the 3d Ising ferromagnet it is clear that the FSS methods employed
in \cite{our1,our4} provide reliable estimates of the multicanonical
parameters more efficiently than the recursion of this paper. On the
other hand, the FSS approach breaks down \cite{our2} for the important
class of disordered systems. Then recursions like the one of this paper
become crucial to enable the method, and the Ising ferromagnet is still
a suitable testing ground to set quantitative performance
scales. These are now given, for the first time, by tables~2 and~4.
Table~4 corresponds to the important case of a typical Edwards--Anderson
Ising spin glass. Future investigations will have to cope with these
standards.  It is my hope that they will bring improvements in
the constant factor, and possibly towards a $V^2$ power law behavior,
which is optimal for any kind of local random walk behavior.
\hfill\break

\section{Appendix}

In this appendix I describe the actually used computer implementation for
the accumulative recursion of the multicanonical parameters. The relevant
Fortran subroutine is listed next. It is not claimed that this subroutine
is in any sense optimal. It just worked sufficiently well for the
described examples.

{\baselineskip 12pt \begin{verbatim}
      SUBROUTINE UPMUCA(IRPT)
C Update of multicanonical parameters.
C HAMUA(*,1):   over-all sum (record keeper only).
C HAMUA(*,2):   LRTRT adjusted over-all sum (record keeper only).
C HAMUA(*,3):   1. weighted sum.
C HAMUA(*,4):   2. weighted sum.
      IMPLICIT REAL*8 (A-H,O-Z)
      IMPLICIT LOGICAL (L)
      PARAMETER (ND=3,NL=08,NS=NL**ND,NRPT=100,NSW=NS)
      PARAMETER (NNH=(ND*NS)/2,NAMIN=NNH/2,FRTRT=3.D0,EPS=1.D-8)
      PARAMETER (HMIN=1.0D00*FLOAT(NS)*FLOAT(NSW))
      COMMON /MEAH/ HA(0:NNH),IAMIN,IAMAX,ITMIN,ITMAX
      COMMON /MUCA/ B(0:NNH),A(0:NNH),HAMU(0:NNH,4),LRTRT(NRPT)
C
      DO IA=ITMIN,ITMAX
      HAMU(IA,1)=HAMU(IA,1)+HA(IA)
      HAMU(IA,2)=HAMU(IA,2)+HA(IA)
      END DO
C
C Retreat strategy (below) implies: range up to IAMAX.GE.ITMAX.
      IAMAM1=IAMAX-1
      DO IA=NAMIN,IAMAM1
      IAP1=IA+1
      HAMIN=MIN(HA(IA),HA(IAP1))
      HAMAX=MAX(HA(IA),HA(IAP1))
      IF(HAMIN.GT.0.5D00)                              THEN
      W1=HAMIN/HAMAX
      HAMU(IA,3)=HAMU(IA,3)+W1*HA(IA)
      HAMU(IA,4)=HAMU(IA,4)+W1*HA(IAP1)*EXP(-4.0D00*B(IAP1))
                                                       END IF
C BETA update (after retreat HAMIN.LE.0.5 possible):
      HAMUMIN=MIN(HAMU(IA,3),HAMU(IA,4))
      IF(HAMUMIN.GT.EPS)                               THEN
      B(IAP1)=-0.25D00*LOG(HAMU(IA,4)/HAMU(IA,3))
                                                       ELSE
      B(IAP1)=B(IA)
                                                       END IF
      END DO
C
C Retreat strategy for hung-up situations:
      LRTRT(IRPT)=.FALSE.
C Besides retreat, update of MUCA A-array is performed
C (range up to IAMAX.GE.ITMAX is needed for this reason).
      DO IA=NAMIN,IAMAM1
      IAP1=IA+1
      IF(HAMU(IAP1,2).GT.HMIN.AND.HAMU(IAP1,2).GT.FRTRT*HAMU(IA,2)) THEN
C The program may need modifications, if there are
C energy values without states in the .LE.IAMAX range.
      IF(HAMU(IA,2).EQ.0) PRINT*,'UPMUCA Warning: IA = ',IA
      IF(.NOT.LRTRT(IRPT)) PRINT*,
     & 'RETREAT! IRPT,IA,HAMUs:',IRPT,IA,HAMU(IAP1,2),HAMU(IA,2)
      LRTRT(IRPT)=.TRUE.
                                                                  END IF
      IF(LRTRT(IRPT))                                THEN
      HAMU(IAP1,2)=HAMU(IAP1,2)/FRTRT
      B(IAP1)=0.0D00
      HAMU(IAP1,3)=HAMU(IAP1,3)/FRTRT
      HAMU(IAP1,4)=HAMU(IAP1,4)/FRTRT
                                                     END IF
      A(IAP1)=A(IA)-4.0D00*(B(IAP1)-B(IA))*FLOAT(IA)
      END DO
C
      IAMAP1=IAMAX+1
      DO IA=IAMAP1,NNH
      B(IA)=B(IA-1)
      A(IA)=A(IA-1)
      END DO
C
      RETURN
      END
\end{verbatim} }

Relevant parameters (to be set) are the dimension {\tt ND}, and lattice size
{\tt NL}. The presented choice is an $8^3$ lattice. {\tt NS} encodes the
lattice size and {\tt NNH} is needed to dimension a number of arrays.

The argument {\tt IRPT} keeps track of the number of repeated calls to
{\tt UPMUCA}. In an outside {\tt DO}--loop {\tt IRPT} runs from 1 to
{\tt NRPT}. Inside our subroutine {\tt NRPT} is only needed to dimension
the {\tt LOGICAL} array {\tt LRTRT}, which keeps track of the number of
``retreats'', to be discussed later. A parameter not needed at all in our
subroutine is {\tt NSW}. It denotes the number of update sweeps performed
in between the calls to {\tt UPMUCA}. In the presented code it is set equal
to the lattice size, corresponding the recursion time $\tau_0^b$ of section~5.

{\tt NAMIN} sets the lower bound on the {\tt IA} range ($I_A$ of section~5)
to which the recursion is applied: {\tt B(IA)=0} for {\tt IA} $\le$
{\tt MAMIN} implements $\beta(I_A)=0$ for $I_A\le N_{\min}$. The other
parameters will be discussed later on.

Most arguments are passed through {\tt COMMON} blocks. On entry the array
{\tt HA} contains the newly assembled statistics, {\it i.e.} the histogram
of the number of times a certain {\tt IA} value (corresponding to an energy
via (23)) has been visited during the last {\tt NSW} sweeps. (The information
is collected after each single spin update. The array {\tt HA} has to be set
to {\tt HA}~$\equiv 0$ after each  call to {\tt UPMUCA}.) Further arguments
passed by the {\tt COMMON} block {\tt MEAH} (measurements) are: {\tt IAMIN},
the smallest {\tt IA} value encountered so far (not used in {\tt UPMUCA});
{\tt IAMAX}, the largest {\tt IA} value encountered so far; {\tt ITMIN}, the
smallest {\tt IA} value encountered during the last {\tt NSW} sweeps, and
{\tt ITMAX}, the largest {\tt IA} value encountered during the last
{\tt NSW} sweeps.

The meaning of the array(s) {\tt HAMU} is explained by the comments at the
beginning of the subroutine.  Central for the code are the lines

{\baselineskip 12pt \begin{verbatim}
      W1=HAMIN/HAMAX
      HAMU(IA,3)=HAMU(IA,3)+W1*HA(IA)
      HAMU(IA,4)=HAMU(IA,4)+W1*HA(IAP1)*EXP(-4.0D00*B(IAP1))
\end{verbatim} }

which implement our equations (19) and (20) recursively. Next, the arrays
{\tt A} and {\tt B} correspond to the multicanonical functions $\beta$
and $\alpha$ the lines

\begin{verbatim}
      B(IAP1)=-0.25D00*LOG(HAMU(IA,4)/HAMU(IA,3))
and
      A(IAP1)=A(IA)-4.0D00*(B(IAP1)-B(IA))*FLOAT(IA)
\end{verbatim}

implement equations (21) and (11). Of course, {\tt A(NAMIN)}$=0$. The
parameter {\tt EPS} prevents that the $\beta$--recursion takes place
without sufficient statistics, and otherwise equation (22) is chosen.

Some complications arise, mainly because a ``retreat'' strategy has been
implemented to get out of certain ``hung--up'' situations. To discuss
them is beyond the scope of this paper, as the relevant (spin glass)
configurations require more detailed investigations first. In short,
an extreme difference between {\tt HAMU(IA+1,2)} and {\tt HAMU(IA,2)} can
turn out to be artificial, such that its statistics is better not trusted.
``Extreme'' is defined by the parameter {\tt FRTRT}, put to 3 in the
presented code. When the thus defined limit is exceeded the assembled
statistics is reduced in weight by the factor 1/{\tt FRTRT} and
$\beta(I_A)$ is put in
the corresponding energy region to $\beta(I_A)=0$ for the next recursion.
The program may thus escape certain traps successfully. However, I like to
remark that one has to chose {\tt FRTRT} to be very large (around 200), if
one likes to calculate multicanonical parameters for
an $24^3$ IF in the range described in section~5.1. The reason is the
peculiar IF density of states anomaly from $I_A=(3N/2)-7$ to
$I_A=(3N/2)-6$ (see table~1). The choice $\beta\equiv 0$ re--creates
an {\tt FRTRT} factor of order $N$.
\hfill\break

{\bf Acknowledgements:} I would like to thank Wolfhard Janke and Claus
Vohwinkel for useful
discussions. The manuscript was partly written at the Institut f\"ur
Physik, Johannes Gutenberg Universit\"at, Mainz. The author likes to
thank Kurt Binder and his group for their hospitality.
\hfill\break 

\section*{Tables}

\begin{table}[h]
\centering
\begin{tabular}{||c|c|c||}                                  \hline
 $E$        & $ I_A $    & $n(I_A)$    \\ \hline
            &            &             \\ \hline
 $-3N$      & $3N/2$     &    2        \\ \hline
 $-3N+4$    &$(3N/2)-1$  &    0        \\ \hline
 $-3N+8$    &$(3N/2)-2$  &    0        \\ \hline
 $-3N+12$   &$(3N/2)-3$  &  $2N$       \\ \hline
 $-3N+16$   &$(3N/2)-4$  &    0        \\ \hline
 $-3N+20$   &$(3N/2)-5$  &  $6N$       \\ \hline
 $-3N+24$   &$(3N/2)-6$  &$2N^2-14N$   \\ \hline
 $-3N+28$   &$(3N/2)-7$  & $30N$       \\ \hline
 $-3N+32$   &$(3N/2)-8$  &$6N^2-66N$   \\ \hline
 $-3N+36$   &$(3N/2)-9$  &$(2N^3-42N^2+1252N)/6$\\ \hline
\end{tabular}
\caption{{\em Finite lattice low temperature expansion for the 3d IF
$(N=L^3)$.}}
\end{table}
\hfill\break

\begin{table}[h]
\centering
\begin{tabular}{||c|c|c|c|c|c|c|c|c||} \hline
 $L$&$n_{\tau}$& $\overline{\tau}$ & $n_{\tau_0^a}$ &
 $\overline{\tau}_0^{\,a}$   &
 $n_{\tau_0^b}$& $\overline{\tau}_0^{\,b}$   &$n_{\tau_1}$&
 $\overline{\tau}_1$       \\ \hline
    &          &              &              &              &
	       &              &              &                \\ \hline
  4 &    548   & 719 (19) E01 & 126          & 661 (41) E02 &
         126   & 557 (46) E02 & 111          & 731 (58) E01   \\ \hline
  6 &    354   & 126 (05) E03 & 252          & 195 (20) E04 &
         126   & 219 (23) E04 & 145          & 129 (10) E03 \\ \hline
  8 &    559   & 881 (23) E03 & 126          & 311 (55) E05 &
         126   & 253 (50) E05 & 125          & 839 (69) E03 \\ \hline
 12 &    322   & 118 (06) E05 & 140          &  95 (32) E07 &
         164   &  90 (15) E07 & 141          & 127 (11) E04 \\ \hline
 16 &    577   & 760 (30) E05 & 180          &              &
           2   & 14 (big) E09 & 180          & 746 (54) E05 \\ \hline
\end{tabular}
\caption{{\em Tunneling and recursion times for the 3d IF.}}
\end{table}
\hfill\break

\begin{table}[h]
\centering
\begin{tabular}{||c|c|c|c|c|c|c|c||} \hline
 $L$& $\overline{\tau}$& $\overline{\tau}_{0.0}$&
		    $\overline{\tau}_{0.5}$& $\overline{\tau}_{1.0}$
                                                             \\ \hline
    &                  &             &             &         \\ \hline
  4 & 398 (15) E02     & 144 E02     & 304 E02     & 411 E03 \\ \hline
  6 & 336 (30) E04     & 436 E03     & 131 E04     & 670 E05 \\ \hline
  8 & 171 (46) E06     & 505 E04     & 282 E05     & 213 E08 \\ \hline
 12 & 139 (77) E08     & 408 E06     & 481 E07     & 544 E08 \\ \hline
\end{tabular}
\caption{{\em Mean $\overline{\tau}$ and some $q$--tiles $\tau_q$ for
the 3d EAI tunneling time.}}
\end{table}
\hfill\break

\begin{table}[h]
\centering
\begin{tabular}{||c|c|c|c|c|c|c|c|c||} \hline
 $L$&$n_{\tau}$& $\overline{\tau}$ & $n_{\tau_0^a}$ &
 $\overline{\tau}_0^{\,a}$   &
 $n_{\tau_0^b}$& $\overline{\tau}_0^{\,b}$   &$n_{\tau_1}$&
 $\overline{\tau}_1$       \\ \hline
    &          &              &              &              &
	       &              &              &                \\ \hline
  4 &    185   & 332 (25) E02 & 126          & 523 (19) E02 &
         126   & 409 (22) E02 & 270          & 228 (14) E02   \\ \hline
  6 &    256   & 181 (12) E04 & 126          & 150 (10) E04 &
         126   & 172 (11) E04 & 357          & 131 (08) E04   \\ \hline
  8 &    134   & 272 (26) E05 & 252          & 245 (13) E05 &
         252   & 253 (16) E05 & 207          & 203 (16) E05   \\ \hline
 12 &          &              &              &              &
          10   & [19 (4) E06] &  13          &  92 (31) E08   \\ \hline
\end{tabular}
\caption{{\em Tunneling and recursion times for typical 3d EAI
spin glass realizations.}}
\end{table}

\begin{thebibliography}{99}
\bibitem{foot1} Note that the recursions given in \cite{our2} and
\cite{Lee} are algebraically identical \cite{our5} in their central parts.
Details, like the energy range to which the recursion is applied, and the
prescription given for values outside this range, may
vary. They are discussed in section~3 of this paper.
\bibitem{foot2} It is straightforward, but tedious, to modify the
subroutine {\tt UPMUCA} of the appendix, such that it can cope with
$n(I_A)=0$ for (isolated) $I_A$ values.
\end{thebibliography}
\end{document}